\documentclass[showkeys,12pt, preprint,preprintnumbers,nofootinbib, groupedaddress,superscriptaddress,amsmath,amssymb]{revtex4}
\usepackage{graphicx}
\usepackage{dcolumn}
\usepackage{bm}
\usepackage{amssymb}
\usepackage{amsmath}
\usepackage{epsfig}    
\usepackage{color}
\usepackage{slashed}
\usepackage{hhline}

\def\be{\begin{equation}}
\def\ee{\end{equation}}
\newcommand{\bea}{\begin{eqnarray}}
\newcommand{\eea}{\end{eqnarray}}
\newcommand{\nn}{\nonumber}

\numberwithin{equation}{section}

\begin{document}

{\begin{flushright}{KIAS-P17007}
\end{flushright}}

\title{ A flavor dependent gauge symmetry, \\
Predictive radiative seesaw and LHCb anomalies}

\author{ P. Ko}
\email{pko@kias.re.kr}
\affiliation{School of Physics, KIAS, Seoul 02455, Korea}
\affiliation{Quantum Universe Center, KIAS, Seoul 02455, Korea}

\author{Takaaki Nomura}
\email{nomura@kias.re.kr}
\affiliation{School of Physics, KIAS, Seoul 02455, Korea}

\author{Hiroshi Okada}
\email{macokada3hiroshi@cts.nthu.edu.tw}
\affiliation{Physics Division, National Center for Theoretical Sciences, Hsinchu, Taiwan 300}

\date{\today}

\begin{abstract}
We propose a predictive radiative seesaw model at one-loop level with a flavor dependent 
gauge symmetry $U(1)_{xB_3-xe-\mu+\tau}$ and Majorana fermion dark matter.   For the neutrino mass matrix, 
we obtain an $A_1$ type  texture (with two zeros) that provides us several predictions such as the 
normal ordering for the neutrino masses. We analyze the constraints from lepton flavor violations, relic density 
of dark matter, and collider physics for the  new $U(1)_{xB_3-xe-\mu+\tau}$ gauge boson. 
Within the allowed region,  
the LHCb anomalies in $B\rightarrow K^* \mu^+ \mu^-$ and  $B\rightarrow K \ell^+ \ell^-$ with $\ell=e$ or $\mu$ can be resolved, 
and such $Z'$ could be also observed at the LHC. 

\end{abstract}
\maketitle
\newpage

\section{Introduction}
Non-zero neutrino masses and their flavor mixings require physics beyond the standard model (SM). 
One of the attractive mechanisms for generating neutrino masses and mixings is the so-called radiative seesaw 
in which the smallness of neutrino mass is explained by the suppression from the loop factor.   
In this class of radiative neutrino mass models, dark matter (DM) candidate often  appears  naturally if we assign
dark $Z_2$ parity to stabilize the DM candidates {(some earlier works are found in refs.~\cite{Ma:2006km, Kajiyama:2013zla,
Krauss:2002px, Aoki:2008av, Gustafsson:2012vj}).}

The predictive neutrino mass model can be achieved by applying some symmetry which distinguishes fermion 
flavor.  Flavor dependent $U(1)$ gauge symmetry is one of the interesting candidates which is discussed in the 
case of tree level neutrino mass generation~\cite{Crivellin:2015lwa,Kownacki:2016pmx}.
Furthermore, flavor dependent $U(1)$ gauge symmetries including the quark sector have been motivated 
in order to explain various anomalies \footnote{Chiral $U(1)'$ gauge theories with additional Higgs doublets
carrying nonzero $U(1)'$ charges were discussed in Refs.~\cite{Ko:2011vd,Ko:2011di,Ko:2012ud,Ko:2012sv} 
in order to accommodate the top forward-backward asymmetry (FBA) at the Tevatron. 
Since this anomaly has been less significant now, we do not consider this case further. But the model building 
issues addressed in Refs.~\cite{Ko:2011vd,Ko:2011di,Ko:2012ud,Ko:2012sv} still remain valid and
relevant in other flavor dependent $U(1)'$ models for $B$ physics anomalies.}
in $B \to K^{(*)} \mu^+ \mu^-$~\cite{Crivellin:2015lwa}; 
$2.6 \sigma$ anomaly in lepton-universality in the ratio $R_K \equiv {\rm BR}(B \to K \mu^+ \mu^-)/{\rm BR}(B \to K \mu^+ \mu^-) = 0.745 ^{+0.090}_{-0.074} \pm 0.036$ by the LHCb \cite{lhcb-2014}, 
and sizable deviation measured in angular distributions of $B \to K^*\mu^+ \mu^-$ \cite{lhcb-2013}. 
These anomalies can be accounted { by a shift in} the Wilson coefficient $C_9$ of the semileptonic operator $O_9$~\cite{Descotes-Genon:2015uva, Hurth:2016fbr} which can be induced by flavor dependent $Z'$ 
interaction in down quark sector.

In this paper, we propose a radiative seesaw model based on  flavor dependent and anomaly free 
$U(1)_{x B_3 - x e +\mu -\tau}$ gauge symmetry and extra discrete $Z_2$ symmetry to ensure DM stability. 
The active neutrino mass matrix is induced at one loop level where $Z_2$ odd particles propagate inside the loop 
including the DM candidate which is the lightest $Z_2$ odd SM singlet Majorana fermion with nonzero $U(1)'$ 
charge.   Then structure of the mass matrix for the Majorana fermion is restricted and determined by the flavor dependent $U(1)'$ charge assignments.  
We also study phenomenology associated with $Z'$ boson, such as collider constraints, signatures at the LHC 
and the Wilson coefficients contributing to $B \to K^{(*)} \mu^+ \mu^-$ obtained from flavor dependent $Z'$ interaction. 
Then we show predictions in the neutrino mass matrix by carrying out numerical analysis taking into account constraints from lepton flavor violation, thermal relic density of DM and various constraints on $Z'$ interaction.

In Sec. II, we introduce our model Lagrangian and discuss particle properties and their interactions. In Sec.III we discuss phenomenology including neutrino mass matrix, charged lepton flavor violation, relic density of DM, and some processes related to $Z'$ gauge boson including the LHCb anomalies. The numerical analysis is carried out in Sec. IV to find out the parameter region satisfying experimental constraints and to obtain some prediction for neutrino physics. Finally we summarize the results in Sec. V.

 \section{Model Lagrangian and particle properties} 

In this section, we introduce our model and discuss some properties for our analysis in the following sections. 
In the fermion sector, we introduce $SU(2)_L$ singlet Majorana fermions $N_{R_{e,\mu,\tau}}$, and impose a flavor dependent gauge symmetry  $U(1)'\equiv U(1)_{xB_3-xe-\mu+\tau}$ as summarized in Table~\ref{tab:1}, where 
$x(\neq1)$ is an arbitrary number~\footnote{
{Notice here that all the components of neutrino mass matrix are nonzero for $x=1$, which originates from the 
structure of the right-handed neutrino mass matrix (see Eq. (II.6) below). It follows from the fact that one cannot 
distinguish $N_{R_e}$ from $N_{R_\mu}$.   Then we would lose predictability on the neutrino sector. 
Therefore we shall choose $x \neq 1$ in this paper and keep predictability on the neutrino sector.}}.

This combination of $U(1)'$ is known as anomaly free of the gauge symmetry
~\cite{Crivellin:2015lwa}.~\footnote{In this reference, the authors provide several possibilities of charge 
assignments, depending on which a different type of prediction can be obtained in the neutrino sector
~\cite{Fritzsch:2011qv}. }
Note here that we ignore the kinetic mixing between $U(1)'$ and $U(1)_Y$ assuming it is negligibly small.
In addition, $Z_2$-odd parity is assigned for the new fermion $N_R$'s in order to forbid the tree 
level neutrino masses or(and) to assure the stability of dark matter (DM).

 \begin{widetext}
\begin{center} 
\begin{table}
\begin{footnotesize}
\begin{tabular}{|c||c|c|c|c|c|c||c|c|c|c|c|c|c|c|c|c|}\hline\hline  
&\multicolumn{6}{c||}{Quarks} & \multicolumn{9}{c|}{Leptons} \\\hline
Fermions ~&~ $Q_L^i$ ~&~ $u_R^i$ ~&~ $d_R^i$ ~&~ $Q_L^3$ ~&~ $b_R$ ~ &~ $t_R$ ~&~ $L_{L_e}$ ~&~ $L_{L_\mu}$ ~&~ $L_{L_\tau}$ ~&~ $e_R$ ~&~ $\mu_R$ ~&~ $\tau_R$ ~&~ $N_{R_e}$ ~&~ $N_{R_\mu}$ ~&~ $N_{R_\tau}$~
\\\hline 
$SU(3)_C$ & $\bm{3}$  & $\bm{3}$  & $\bm{3}$ & $\bm{3}$ & $\bm{3}$ &
 $\bm{3}$  & $\bm{1}$  & $\bm{1}$  & $\bm{1}$   & $\bm{1}$  & $\bm{1} $  & $\bm{1}$ & $\bm{1}$  & $\bm{1} $  & $\bm{1}$ \\\hline 
 $SU(2)_L$ & $\bm{2}$  & $\bm{1}$  & $\bm{1}$ & $\bm{2}$ & $\bm{1}$ &
 $\bm{1}$  & $\bm{2}$  & $\bm{2}$  & $\bm{2}$   & $\bm{1}$  & $\bm{1}$   & $\bm{1}$  & $\bm{1}$  & $\bm{1}$   & $\bm{1}$ \\\hline 
$U(1)_Y$ & $\frac16$ & $\frac23$  & $-\frac{1}{3}$ & $\frac16$ & $-\frac{1}{3}$  & $\frac{2}{3}$ & $-\frac{1}{2}$
 & $-\frac12$ & $-\frac12$  & $-1$ &  $-1$  &  $-1$  & $0$ &  $0$  &  $0$ \\\hline
 $U(1)'$ & $0$ & $0$  & $0$ & $\frac{x}{3}$ & $\frac{x}{3}$  & $\frac{x}{3}$ & $-x$  & $-1$ & $1$ & $-x$  & $-1$   & $1$ & $-x$  & $-1$   & $1$ \\\hline
$Z_2$ & $+$ & $+$  & $+$ & $+$ & $+$
& $+$ & $+$  & $+$ & $+$ & $+$ & $+$ & $+$& $-$ & $-$ & $-$ \\\hline
\end{tabular}
\caption{Field contents of fermions
and their charge assignments under $SU(2)_L\times U(1)_Y\times U(1)'\times Z_2$, where $U(1)'\equiv U(1)_{xB_3-xe-\mu+\tau}$ ($x\neq1$), and each of the flavor index is defined as $i=1,2$.}
\label{tab:1}
 \end{footnotesize}
\end{table}
\end{center}
\end{widetext}
\begin{table}[t]
\centering {\fontsize{10}{12}
\begin{tabular}{|c||c|c|c|c|c||c|c|}\hline\hline
&\multicolumn{5}{c||}{VEV$\neq 0$} & \multicolumn{1}{c|}{Inert } \\\hline
  Bosons  &~ $\Phi_1$   ~& ~ $\Phi_2$ ~ &~ $\varphi_{x/3}$ ~ &~ $\varphi_2$     ~ &~ $\varphi_{1-x}$   ~ &~ $\eta$ ~ \\\hline
$SU(2)_L$ & $\bm{2}$ & $\bm{2}$ & $\bm{1}$  & $\bm{1}$   & $\bm{1}$ & $\bm{2}$ \\\hline 
$U(1)_Y$ & $\frac12$ & $\frac12$ & $0$  & $0$ & $0$ & { $\frac12$ }  \\\hline
 $U(1)'$ & $0$ & $-\frac{x}{3}$ & $\frac{x}{3}$ & $2$ & $1-x$ & $0$   \\\hline
$Z_2$ & $+$ & $+$ & $+$ & $+$ & $+$ & $-$ \\\hline
\end{tabular}%
} 
\caption{Field contents of scalar bosons
and their charge assignments under $SU(2)_L\times U(1)_Y\times U(1)'\times Z_2$. }
\label{tab:2}
\end{table}

In the scalar sector, we introduce an $SU(2)_L$ doublet inert scalar field $\eta$, new Higgs doublet $\Phi_2$ 
with extra $U(1)$ charge, and  three $SU(2)_L$ singlet  scalars $\{\varphi_2, \varphi_{1-x}, \varphi_{x/3}\}$, 
where the lower indices represent their charges under $U(1)'$  as summarized in Table~\ref{tab:2}.  
We assume that two Higgs doublets $\Phi_1 , \Phi_2$ 
and $SU(2)$ singlet fields $\{ \varphi_2, \varphi_{1-x}, \varphi_{x/3} \}$ respectively break electroweak and $U(1)'$ gauge 
symmetries spontaneously by their nonzero vacuum expectation values (VEVs), which are 
denoted by $v/\sqrt2$, $v'/\sqrt2$, $v_2/\sqrt2$, $v_{1-x}/\sqrt2$ and $v_{x/3}/\sqrt2$.
The new Higgs doublet $\Phi_2$ is introduced in order to induce quark mass matrix element 
which mix the 3rd generation with first and second generations. 

The Higgs potential of two doublets are written by
\begin{align}
V \supset & \mu_1^2 \phi_1^\dagger \phi_1 + \mu_1^2 \phi_1^\dagger \phi_1 + \lambda_1 (\Phi_1^\dagger \Phi_1)^2 + \lambda_2 (\Phi_2^\dagger \Phi_2)^2 \nonumber \\
&+ \lambda_3 (\Phi_1^\dagger \Phi_1) (\Phi_2^\dagger \Phi_2) + \lambda_4 |\Phi_1^\dagger \Phi_2|^2 + \mu \varphi_{x/3} \Phi_1^\dagger \Phi_2
\end{align}
where $\varphi_{x/3}$ provides a dim-3 operator. 
Note that we have a massless Goldstone boson associated with second Higgs doublet without the dim-3 operator.
Thus $\varphi_{x/3}$ allows us to avoid the constraints of a massless boson from SU(2) doublet scalar.
Note also that scalar potential of $\varphi_2$ and $\varphi_{1-x}$ has global symmetries which would induce a massless Goldstone boson since the potential is given by $|\varphi_2|^2$ and $|\varphi_{1-x}|^2$ due to the 
$U(1)'$ symmetry. Such global symmetries can be avoided by introducing $U(1)'$-charged scalar; for example 
$\varphi_{x-3}$ with $U(1)'$ charge $(x-3)$ provides a term $\varphi_{x-3} \varphi_{1-x} \varphi_2$ which violate 
dangerous global symmetries. In this paper, we assume all scalar bosons have non-zero masses and we 
abbreviate the complete analysis of the scalar potential. 

{\it Yukawa interactions}:
Under these fields and symmetries, the renormalizable Lagrangians for quark and lepton sector are given by 
\begin{align}
-{\cal L}_{Q} = & (y_u)_{i j}\bar Q_{L_i}  \tilde\Phi_1 u_{R_j} + (y_d)_{ij} \bar Q_{L_i}\Phi_1 d_{R_j} + (y_u)_{33}\bar Q_{L_3}  \tilde\Phi_1 t_R + (y_d)_{33} \bar Q_{L_3}\Phi_1 b_{R} \nn \\
& + (\tilde y_u)_{3i}\bar Q_{L_3}  \tilde\Phi_2 u_{R_i} + (\tilde y_d)_{i3} \bar Q_{L_i} \Phi_2 b_{R}  +{\rm h.c.}, \label{eq:lag-quark}\\
-{\cal L}_{L} = & y_e \bar L_{L_e}  \Phi_1 e_R  +y_\mu \bar L_{L_\mu}  \Phi_1 \mu_R 
+y_\tau \bar L_{L_\tau} \Phi_1 \tau_{R} 
+y_{N_1} \bar L_{L_e} \tilde\eta N_{R_e}  + y_{N_2} \bar L_{L_\mu} \tilde\eta N_{R_\mu} + y_{N_3} \bar L_{L_\tau} \tilde\eta N_{R_\tau} 
\nn\\&
+ M_{23}(\bar N_{R_\mu}^c N_{R_\tau} + \bar N_{R_\tau}^c N_{R_\mu})
+ f_1 \varphi_2 \bar N_{R_\mu}^c N_{R_\mu} + f_2 \varphi_2^* \bar N_{R_\tau}^c N_{R_\tau}
\nn\\&
+ f_{13} \varphi_{1-x}^*(\bar N_{R_e}^c N_{R_\tau} +\bar N_{R_\tau}^c N_{R_e})
+ {\rm c.c.},
\label{eq:lag-lepton}
\end{align}
where $(i,j)=1,2$, $\tilde\Phi\equiv i\sigma_2\Phi^*$, and  $\sigma_2$ is the second Pauli matrix.

After two Higgs doublet develops nonzero VEVs, we obtain the quark mass matrix such that 
\begin{align}
\label{eq:Mu}
M^u &= \frac{1}{\sqrt{2}} \left( \begin{array}{ccc} v (y_u)_{11} & v (y_u)_{12} & 0 \\ v (y_u)_{21} & v (y_u)_{22} & 0 \\ 0 & 0 & v (y_u)_{33} \end{array} \right) 
+ \left( \begin{array}{ccc} 0 & 0 & 0 \\ 0 & 0 & 0 \\  (\xi_u)_{31} & (\xi_u)_{32} & 0 \end{array} \right), \\ 
\label{eq:Md}
M^d &= \frac{1}{\sqrt{2}} \left( \begin{array}{ccc} v (y_d)_{11} & v (y_d)_{12} & 0 \\ v (y_d)_{21} & v (y_d)_{22} & 0 \\ 0 & 0 & v (y_d)_{33} \end{array} \right)
+ \left( \begin{array}{ccc} 0 & 0 & (\xi_d)_{13} \\ 0 & 0 & (\xi_d)_{23} \\ 0 & 0 & 0 \end{array} \right),
\end{align}
where $\xi_{u,d} \equiv \tilde y_{u,d} v'/\sqrt{2}$.
Note that the second term of Eqs.~(\ref{eq:Mu}) and (\ref{eq:Md}) are obtained from the last two terms of Eq.~(\ref{eq:lag-quark}) associated with the VEV of second Higgs $\Phi_2$.
Thus the mass matrices have the same structure as discussed in Ref.~\cite{Crivellin:2015lwa}.
Note that elements with $\xi_{u,d}$ are considered to be small perturbation effects 
generating realistic $3 \times 3$ CKM mixing matrix, and the  $(33)$  elements are 
$v (y_{u(d)})_{33} \sim \sqrt{2} m_{t(b)}$.
As in the SM, the quark mass matrices are diagonalized by unitary matrices $U_{L, R}$ and $D_{L,R}$ which 
change quark fields from interaction basis to mass basis: 
$u_{L,R} \to U_{L,R}^\dagger u_{L,R} (d_{L,R} \to D_{L,R}^\dagger d_{L,R})$. 
Thus the mass matrices $M^{u,d}$ are related to diagonal mass matrices as follows:
\begin{equation}
M^d = D_L m^d_{\rm diag} D_R^\dagger, \quad M^u = U_L m^u_{\rm diag} U_R^\dagger,
\end{equation}
where $m^d_{\rm diag} = {\rm diag}(m_d,m_s,m_b)$ and $m^u_{\rm diag} = {\rm diag}(m_u,m_c,m_t)$.
We find that off-diagonal elements associated with 3rd generations are more suppressed for $M^u (M^u)^\dagger$ and $(M^d)^\dagger M^d$ than those in $(M^u)^\dagger M^u$ and $M^d (M^d)^\dagger$. 
Then $U_L$ and $D_R$ can be approximated to be close to unity matrix since they are respectively associated with diagonalizaition of $M^u (M^u)^\dagger$ and $(M^d)^\dagger M^d$. 
Thus we can approximate $V_{CKM} = U^\dagger_L D_L \simeq D_L$ and $D_R \simeq {\bm 1}$~\cite{Crivellin:2015lwa}.
The details of quark Yukawa couplings with two Higgs doublet are discussed in Ref.~\cite{Crivellin:2015lwa}, and we omit the further discussion here.


{\it $Z'$ couplings to SM fermions}:   After the aforementioned fields rotations into the mass basis, 
the $Z'$ couplings to the SM fermions are written as 
\begin{align}
{\cal L}_{Z'ff} \supset & g' \left(-x \bar e \gamma^\mu e - \bar \mu \gamma^\mu \mu + \bar \tau \gamma^\mu \tau -x \bar \nu_e \gamma^\mu P_L \nu_e - \bar \nu_\mu \gamma^\mu P_L \nu_\mu + \bar \nu_\tau \gamma^\mu P_L \nu_\tau + \frac{x}{3} \bar t \gamma^\mu t \right) Z'_\mu \nonumber \\
&+ x g' \left( \bar d_\alpha \gamma^\mu P_L d_\beta \Gamma^{d_L}_{\alpha \beta} + \bar d_\alpha \gamma^\mu P_R d_\beta \Gamma^{d_R}_{\alpha \beta} \right) Z'_\mu ,
\label{eq:int_Z'}
\end{align}
where $g'$ is the gauge coupling constant associated with the $U(1)'$. 
The coupling matrices $\Gamma^{d_R}$ and $\Gamma^{d_L}$ for down-type quarks are given approximately by
\begin{equation}
\Gamma^{d_L} \simeq \frac{1}{3} \left( \begin{array}{ccc} |V_{td}|^2 & V_{ts}V^*_{td} & V_{tb} V^*_{td} \\ V_{td} V^*_{ts} & |V_{ts}|^2 & V_{tb} V^*_{ts} \\ V_{td} V^*_{tb} & V_{ts} V^*_{tb} & |V_{tb}|^2  \end{array} \right), \quad 
\Gamma^{d_R} \simeq  \left( \begin{array}{ccc} 0 & 0 & 0 \\ 0 & 0 & 0 \\ 0 & 0 & \frac{1}{3} \end{array} \right),
\label{eq:ckm}
\end{equation}
where $V_{qq'}$s are the elements of CKM matrix and we applied the relation $V_{CKM} \simeq D_{ L}$ 
as we discussed above.
\\
{\it Exotic Majorana fermion mass matrix} is defined in the basis $[N_{R_e},N_{R_\mu},N_{R_\tau}]^T$ as follows:
\begin{align}
M_N\equiv \left[\begin{array}{ccc}  0 & 0 & M_{13}  \\ 0 & M_{22} & M_{23} \\ M_{13} & M_{23} & M_{33} \end{array}\right],\label{eq:MN}
\end{align}
where we simply assume these elements are positive and real, and define $M_{22}\equiv f_1 v_2/\sqrt2$, $M_{33}\equiv f_2 v_2/\sqrt2$, and $M_{13}\equiv f_{13}v_{1-x}/\sqrt2$. Then $M_N$ is diagonalized by orthogonal mixing matrix $V$ as 
\begin{align}
V^T M_N V =D_N\equiv \left[M_1,M_2,M_3\right],\quad N_{R_{e,\mu,\tau}}=V N_{R_{1,2,3}},\label{eq:N-mix}
\end{align}
where $M_{1,2,3}$ is the mass eigenstate.

\section{Phenomenology}
\subsection{Active neutrino masses and lepton flavor violating processes}

{\it The Active neutrino mass matrix} is then given at  one-loop level by~\cite{Ma:2006km}
\begin{align}
-(m_\nu)_{ij}&= \frac1{32\pi^2} \sum_{k=1}^3 (y_{N_i} V_{ik}) D_{N_k} (y_{N_j} V_{jk})
\left(\frac{m_R^2}{m_R^2- D_{N_k}^2} \ln\left[\frac{m_R^2}{D_{N_k}^2}\right] - \frac{m_I^2}{m_R^2- D_{N_k}^2} \ln\left[\frac{m_I^2}{D_{N_k}^2}\right] \right)\nn\\
&\approx \frac1{8\pi^2} \frac{\lambda_5 v^2}{m_R^2+m_I^2} \sum_{k=1}^3 y_{N_i} (V_{ik} D_{N_k}V_{kj}^T) y_{N_j} 
= \frac1{8\pi^2} \frac{\lambda_5 v^2}{m_R^2+m_I^2}  y_{N_i} (M_N)_{ij} y_{N_j},
\label{eq:neut-mass}
\end{align}
where $\lambda_5$ is the quartic coupling of $(\Phi^\dag\eta)^2$, $m_{R(I)}$ is the mass eigenstate of 
real(imaginary) part of neutral component of $\eta$, and we have used Eq.(\ref{eq:N-mix}) in the last equation. 
Here we assume to be $D_N<<m_{R(I)}$, which could be natural if we consider the fermion DM case.
Since $y_N$ is diagonal, the form of active neutrino mass matrix is proportional to the one of $M_N$ in Eq.(\ref{eq:MN}), therefore we have some predictions of type $A_1$  through the texture analysis~\cite{Fritzsch:2011qv}.
Then $M_N$ can be rewritten in terms of PMNS matrix $U_{MNS}$ and mass eigenvalues of active neutrino $D_\nu$ by
$m_\nu\equiv U_{MNS} D_\nu U_{MNS}^T$, where we define $D_\nu\equiv U_{MNS} ^\dag m_\nu U_{MNS}^*$.
Combining Eq.~(\ref{eq:neut-mass}), $D_N$ can be rewritten in terms of neutrino observables and some input parameters such as $y_N$ by
\begin{align}
D_N\approx -\epsilon V^* y_N^{-1} U_{MNS} D_\nu U_{MNS}^T y_N^{-1} V^\dag,
\end{align}
where $\epsilon\equiv \frac{8\pi^2(m_R^2+m_I^2)}{\lambda_5 v^2}$.
In our numerical analysis,
we will show some predictions combined with the other phenomenologies such as LFVs and DM,
adapting the recent global data~\cite{Forero:2014bxa} up to 3$\sigma$ confidential level.\\

{\it Lepton flavor violations(LFVs)} are induced from the term $y_N$ at one-loop level, and its branching ratio is 
given by
\begin{align}
& {\rm BR}(\ell_i\to\ell_j\gamma)= \frac{48\pi^3\alpha_{\rm em} C_{ij} }{G_F^2 m_{\ell_i}^2}\left(|a_{R_{ij}}|^2+|a_{L_{ij}}|^2\right),\\
& a_{R_{ij}}=\sum_{\alpha=1,2,3}\frac{y^*_{N_i} y_{N_j} V^\dag_{\alpha i} V_{j\alpha} m_{\ell_i}}{(4\pi)^2} F_{lfv}(N_\alpha,\eta^\pm), \\ 
& a_{L_{ij}}=\sum_{\alpha=1,2,3}\frac{y^*_{N_i} y_{N_j} V^\dag_{\alpha i} V_{j\alpha} m_{\ell_j}}{(4\pi)^2} F_{lfv}(N_\alpha,\eta^\pm),  \\
& F_{lfv}(a,b)=\frac{2 m_a^6 +3 m_a^3 m_b^3 -6 m_a^2 m_b^4 +6 m_b^6+12 m_a^4 m_b^2 \ln(m_b/m_a)}{12(m_a^2-m_b^2)^4},
\end{align}
where $\eta^\pm$ is the singly charged component of $\eta$, $G_F \approx 1.17\times10^{-5}$[GeV]$^{-2}$ is the Fermi constant, $\alpha_{\rm em}\approx1/137$ is the fine structure constant, $C_{21}\approx1$, $C_{31}\approx 0.1784$,  and $C_{32}\approx 0.1736$.
Experimental upper bounds are respectively given by ${\rm BR}(\mu\to e\gamma)\lesssim 4.2\times10^{-13}$~\cite{TheMEG:2016wtm}, ${\rm BR}(\tau\to e\gamma)\lesssim 3.3\times10^{-8}$, and ${\rm BR}(\tau\to \mu\gamma)\lesssim 4.4\times10^{-8}$~\cite{Adam:2013mnn}.
\\
{\it Muon anomalous magnetic moment (muon g-2: $\Delta a_\mu$)} can be induced via $y_N$ with negative contribution, which is in conflict with the current experiment $\Delta a_\mu = (26.1 \pm 8.0) \times 10^{-10}$~{\cite{bennett}}.
However another source via the additional $Z'$ gauge boson can also be induced by
\begin{align}
\Delta a_{\mu}^{Z'}\approx \frac{g_{Z'}^2}{8\pi^2}\int_0^1 da \frac{2 r a (1-a)^2}{r(1-a)^2+a}, 
\end{align}
where $r\equiv(m_\mu/M_{Z'})^2$, and $Z'$ is the new gauge vector boson. Thus we could explain the sizable 
muon $(g-2)$ if we can satisfy the constraint from the neutrino trident process: 
$M_{Z'}\lesssim $0.4 GeV with $g'\lesssim10^{-3}$~\cite{Altmannshofer:2014pba}. This can be realized by the limit $x=0$. However this is nothing but a typical gauged $\mu-\tau$ symmetry~\cite{Baek:2015mna}.
Thus we discuss parameter region with heavier $Z'$ mass which does not include the region solving muon $g-2$ in $1 \sigma$ level.  
When we apply the upper bound of $g'/m_{Z'} \lesssim (550 \ {\rm GeV})^{-1}$ from the neutrino trident process~\cite{Altmannshofer:2014pba}, we obtain $\Delta a_\mu \lesssim 3 \times 10^{-10}$, which is 
smaller than the measured value but it is within 3 $\sigma$ level deviation. It could be tested in future 
experiments.  

\subsection{Dark matter}

Here we consider the lightest Majorana fermions $X\equiv N_1$ as our DM, and assume $M_{Z'} > m_X$ 
to forbid the mode of $2X\to 2Z'$ for simplicity \footnote{See Ref.\cite{Ko:2016sxg} in the case of $M_{Z'} < m_X$.}.
Also we neglect  mixings 
among neutral component of $(\Phi,\varphi_2,\varphi_{1-x},\varphi_{x/3})$ to simply suppress Higgs portal interaction for avoiding the constraint from direct detection 
searches. Therefore the dominant contribution to DM annihilation in estimating the relic density arises from Yukawa coupling.

Then the relevant Lagrangian in terms of mass eigenstates is given by
\begin{align}
{\cal L}&=
\sum_{i=1}^3 y_{N_i} V_{i1} [-\bar \ell_i\eta^-P_R X+\bar \nu_i\eta^* P_R X]+{\rm c.c.}
\\&
+\frac{g'}{2}(-x|V_{11}|^2-|V_{21}|^2+|V_{31}|^2)\bar X\gamma^\mu X Z'_\mu+
 {\cal L}_{Z'ff},\nn
\end{align}
where the last term is given in Eq.~(\ref{eq:int_Z'}).
We have three relevant precesses to explain the relic density: $X\bar X\to\ell_i\bar\ell_j$, $X\bar X\to\nu_i\bar\nu_j$, and $X\bar X\to t\bar t$ via the Yukawa terms $y_N$ and the gauge interaction with $Z'$ involving $g'$,
where we have $s,t,u$ channels only for $i= j$, while $t,u$ channels for $i\neq  j$.~\footnote{Since these formulae are complicated, we will include the numerical form instead writing down explicitly.}
We apply the $v_{rel}$ expansion approximation~\cite{Griest:1990kh} to estimate the relic density of DM, 
taking up to the $S$- and $P$-wave contributions in the annihilation amplitudes. 
Then the formula for thermal relic density $\Omega h^2$ is approximately given by~\cite{Srednicki:1988ce}
\begin{align}
\Omega h^2\approx \frac{4.28\times 10^9 x_f^2}{\sqrt{g_*} M_P[(-3+4 x_f) a_{\rm eff}+12 x_f b_{\rm eff}]},
\end{align}
where $M_P\approx 1.22\times10^{19}$[GeV] is the Planck mass, $g_*\approx 100$ is  the total number of effective relativistic degrees of freedom at the time of freeze-out, and $x_f\approx25$ is defined by $M_X/T_f$ at the freeze 
out temperature ($T_f$), $a_{\rm eff}$ is the total contributions to the $S$-wave, and $b_{\rm eff}$ is the total 
contributions to the $P$-wave, respectively.
The observed relic density reported by Planck suggests that $\Omega h^2\approx 0.12$~\cite{Ade:2013zuv}.
But in our numerical analysis below, we will use more relaxed value $0.11\lesssim \Omega h^2\lesssim 0.13$.\\
%
\subsection{$Z'$ phenomenology and experimental constraints on its couplings}
Here we discuss phenomenology of $Z'$ boson such as the constraints on interactions, the contribution to 
$B \to K^{(*)} \ell^+ \ell^-$, and the direct $Z'$ production cross section at the LHC.

{\it LEP constraint:} The $Z'$ couplings to leptons induce the following effective interactions;
\begin{equation}
L_{eff} = \frac{1}{1+\delta_{e \ell}} \frac{g'^2}{M_{Z'}^2} C_{\ell} (\bar e \gamma^\mu e)( \bar \ell \gamma_\mu \ell)
\end{equation}
where $C_{e} = x^2$, $C_\mu = -x$ and $C_\tau = x$ in our charge assignments.
In this case, the strongest constraint comes from the $e^+ e^- \to \mu^+ \mu^-$ measurement at LEP~\cite{Schael:2013ita}: 
\begin{align}
\frac{M_{Z'}}{\sqrt{x}g'}   >  4.6\ {\rm TeV}.\label{eq:lep}
\end{align}
We will impose this constraint in the following numerical analysis. 

{\it The constraint from neutrino trident production}: The couplings of $Z'$ to the second generation of lepton is 
constrained by the neutrino trident process $\nu_e N \to \nu_e N \mu^+ \mu^-$ where $N$ denotes a nucleon. 
Taking into account the CCFR data,  this constraint is roughly approximated as $m_{Z'}/g' \geq 550$ GeV at 
the 95$\%$ C.L. for a heavy $Z'$ boson case~{\cite{Altmannshofer:2014pba}}.
When we take $g' = g_2 (\simeq 0.65)$, the mass of $Z'$ should satisfy $m_{Z'} \geq 358$ GeV.  

{\it $Z'$ contribution to the $b \to s \bar \ell \ell$ decay} : The anomalies in the angular observable $P'_5$ 
associated with full angular distribution of $B \to K^* \mu^+ \mu^-$ (with $K^* \to K^- \pi^+$) and in the 
lepton-universality violation $R_K = {\rm BR}(B \to K \mu^+ \mu^-)/{\rm BR}(B \to K e^+e^-)$ can be accounted by the shift 
in the Wilson Coefficient $C_9^{\mu \mu}$, which is defined by $\Delta B = 1$ effective Hamiltonian as
\begin{equation}
{\cal H}_{\rm eff} = -\frac{G_F \alpha V_{tb} V^*_{ts}}{\sqrt{2}\pi} C_9^{\ell \ell} (\bar s \gamma^\mu P_L b )(\bar \ell \gamma_\mu \ell) + h.c.
\end{equation}
We have suppressed other operators for simplicity, since they do not play any important role regarding those 
two $B$ physics anomalies considered here as long as the Wilson coefficients of those operators do not receive
new physics contributions. 
The global fit of the value for $C_9^{\mu \mu}$ is obtained in Ref.~\cite{Hurth:2016fbr} based on LHCb data 
as follows;
\begin{equation}
\frac{ \Delta C_9^{\mu \mu}}{C_9^{SM}} = -0.21 : ({\rm best\ fit \ value}), \qquad  [-0.27,-0.13]\;({\rm at} \ 1\sigma), \qquad
  [-0.32,-0.08] \; ({\rm at} \ 2\sigma). \label{eq:C9_fit}
\end{equation}
where $\Delta C_9^{\mu \mu}$ indicates new physics contribution and $C_9^{SM} = 4.07$ at $\mu_b = 4.8$ GeV.
Note that the SM contribution $C_9^{SM} $ is lepton flavor universal, unlike to $\Delta C_9^{\mu \mu}$.

In the model proposed in Sec.  II, the flavor-dependent $Z'$ interaction shall induce the following 
effective Hamiltonian: 
\begin{equation}
\Delta {\cal H}_{\rm eff} = \frac{g'^2 V_{tb} V_{ts}^*}{3 M_{Z'}^2} X_\ell (\bar s \gamma^\mu P_L b )(\bar \ell \gamma_\mu \ell)
\end{equation}
where $X_e = x^2$ and $X_\mu = -X_\tau = x$. Thus the shift of $C_9^{\mu \mu}$ relative to its SM value 
would be given by 
\begin{equation}
\Delta C_9^{\mu \mu} = - x \frac{\sqrt{2} \pi}{3 G_F \alpha} \left( \frac{g'}{M_{Z'}} \right)^2.
\end{equation}
Therefore, applying the LEP constraint Eq.~(\ref{eq:lep}), we find the range of $\Delta C_9^{\mu \mu}$ such that
\begin{equation}
\label{eq:C9mm}
-0.46 \lesssim \Delta C_9^{\mu \mu} \leq 0.
\end{equation}
where the dependence on $x$ is canceled since the upper limit of $g'/M_{Z'}$ is proportional to $1/\sqrt{x}$.
The magnitude of $|\Delta C_9^{\mu \mu}|$ is smaller than best fit value 
($\Delta C_9^{\mu \mu} \simeq -0.85$) but it is within $2 \sigma$ range as shown in Eq.~(\ref{eq:C9_fit}).

Note that $\Delta C_9^{ee}$ is suppressed by an extra factor of $x$ in our model. 
Thus it is possible to explain the anomaly in lepton-universality in $b\rightarrow s \bar \ell \ell$:  
$R_K = {\rm BR}(B \to K \mu^+\mu^-)/{\rm BR}(B \to K e^+e^-)=0.745^{+0.090}_{-0.074} \pm 0.036$ measured 
by LHCb, which shows a $2.6\sigma$ deviation from the SM prediction. 
Here the $R_K$ can be rewritten in terms of $X^{\ell \ell} =\Delta C^{\ell \ell}_9 - \Delta C^{\ell \ell}_{10}$ ($\ell=e,\mu$) where $\Delta C_{10}^{\ell \ell} =0$ in our model, and its allowed region is found to be~\cite{Hiller:2003js, Hiller:2014yaa};
$ 0.7 \leq Re[X^e - X^\mu] \leq 1.5$,
applying the $R_K$ data with $1\sigma$ errors. This condition can be interpreted as 
\begin{align}
 -0.75 \lesssim \Delta C_9^{\mu \mu} \lesssim -0.35,\label{eq:rk-constraint}
\end{align}
where $X^e << X^\mu$ is used. Therefore our value of $C_9^{\mu \mu}$ in Eq.~(\ref{eq:C9mm}) can be accommodated with the range.

\begin{figure}[t]
\begin{center}
\includegraphics[width=70mm]{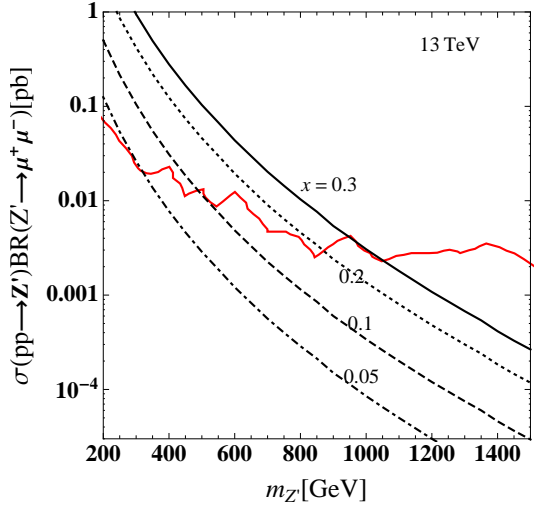} 
\caption{$\sigma(pp \to Z')BR(Z' \to \mu^+ \mu^-)$ as a function of $m_{Z'}$ at $\sqrt{s} = 13$ TeV with various values of $x$. The red curve shows the upper limit from the ATLAS experiment.
}   \label{fig:ZpLHC}
\end{center}\end{figure}

{\it $Z'$ production at the LHC} : 
The $U(1)'$ gauge boson $Z'$ can be produced at the LHC since it couples to quarks.
The dominant production process is given by $\bar b b \to Z'$ where the couplings to other quarks are 
suppressed by small CKM matrix elements (see Eq.~(\ref{eq:ckm})).
The $Z'$ mainly decays into $\mu^+ \mu^-$ and $\tau^+ \tau^-$ pairs with their branching ratios (BR's) as 
${\rm BR}(Z' \to \mu^+ \mu^-) \simeq {\rm BR}(Z' \to \tau^+ \tau^-) \simeq 0.5$ for small $x(\lesssim 0.3)$ where 
we assume masses of scalar bosons with couplings which is not suppressed by $x$ are heavier than $M_{Z'}/2$.
Thus the dimuon channel provides the most clear signature of $Z'$.
To estimate the production cross section for $pp \to Z' \to \mu^+ \mu^-$, we implement the relevant interactions 
into CalcHEP~\cite{Belyaev:2012qa} and use the CTEQ6 parton distribution functions (PDFs)
~\cite{Nadolsky:2008zw}.
Fig.~\ref{fig:ZpLHC} shows the $\sigma(pp \to Z'){\rm BR}(Z' \to \mu^+ \mu^-)$ at $\sqrt{s} = 13$ TeV as 
a function of $m_{Z'}$ where we have fixed $g' = g_2$ and applied various values of $x$.
The cross section is compared with the upper bound from the ATLAS experiments which is indicated as red curve~\cite{Aaboud:2016cth}.
We find that $m_{Z'} < 1$ TeV is allowed for $x < 0.3$ and the constraint is weaker for smaller value of $x$.
Further parameter region can be tested by searching for the dimuon signature of $Z'$ at the LHC run 2. 
Here $pp \to \mu^+ \mu^-$ process in the SM provides a background of the signal events and 
the cross section is $\sigma \sim 0.1$ pb when we apply invariant mass cuts of $M_{\mu^+ \mu^-}> $ 
400 GeV. Thus sizable significance can be obtained with sufficient integrated luminosity; for example 
significance of $N_{\rm signal}/\sqrt{N_{\rm BG}} \sim 3$ is obtained with 100 fb$^{-1}$ when the signal 
cross section is $0.003$ pb, where $N_{\rm signal (BG)}$ is the number of signal(background) events. 
The significance can be further improved by taking appropriate kinematical cuts, however, the detailed 
event simulation is beyond the scope of this work.

\section{Numerical analysis}
\begin{figure}[t]
\begin{center}
\includegraphics[width=70mm]{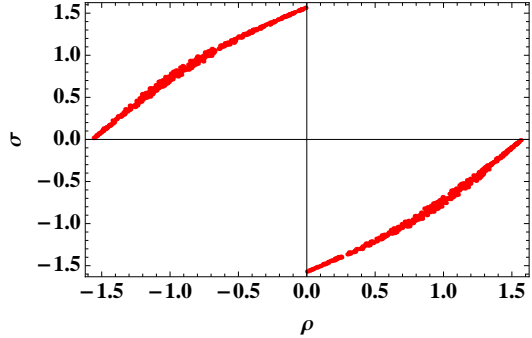} 
\caption{Correlation between $\rho$ and $\sigma$ which are Majorana phases. 
}   \label{fig:majo-p}
\end{center}\end{figure}

In this section, we perform the numerical analysis and show some predictions.
First of all, we select the range of input parameters as follows: 
\begin{align}
& x\in[0.001,0.5],\ y_{N_1}  \in[10^{-5},10^{-4}],\  y_{N_2}  \in[10^{-4},10^{-3}],\  y_{N_1}  \in[10^{-3},10^{-2}],\
\delta_{CP}\in[0,2\pi],\nn\\
& m_{\nu_3}\in [10^{-12},10^{-9}]\ [{\rm GeV}], \quad m_{R}\approx m_I \in [3000,5000]\ [{\rm GeV}], \quad 
M_{Z'}\in [100,1500]\ [{\rm GeV}],
\end{align}
where $\delta_{CP}$ is Dirac phase in the neutrino sector, and we fix the new $U(1)'$ gauge coupling to be $g'=g_2(\approx0.654)$.
Due to the type $A_1$ texture of the neutrino mass matrix, obvious predictions are as follows, 
which are independent of the other phenomenologies as well as the above ranges of input parameters 
as already discussed in ref.~\cite{Fritzsch:2011qv}.
\begin{enumerate}
\item {\it The third neutrino mass is almost fixed to be $m_{\nu_3}\approx (4.8-5.3)\times 10^{-11}$ {\rm [GeV]}.}
\item {\it Only the normal ordering of the neutrino masses is allowed.} 
\item {\it Two Majorana CP phases $\rho$ and $\sigma$ correlate each other and behave as the red line in Fig.~\ref{fig:majo-p},  where $V\equiv U_{MNS} P$ with $P\equiv{\rm diag.(e^{i\rho},e^{i\sigma},0)}$ in 
Ref.~\cite{Fritzsch:2011qv}.   And ${\rm sign}(\sigma  \rho ) < 0$ is predicted.}
\item {\it Neutrinoless double beta decay is predicted to be $\langle m_{ee}\rangle\equiv \sum_{i=1-3}m_{\nu_i} V_{ei}^2\approx {\cal O}$(0.01) eV, which follows from the above two predictions.} 
\end{enumerate}
Here we have used the global neutrino oscillation data at 3$\sigma$ confidential level~\cite{Forero:2014bxa}.
Notice here that  $\delta_{CP}$ is allowed in all the possible range, $\delta_{CP}\in [0,2\pi] $.

\begin{figure}[t]
\begin{center}
\includegraphics[width=70mm]{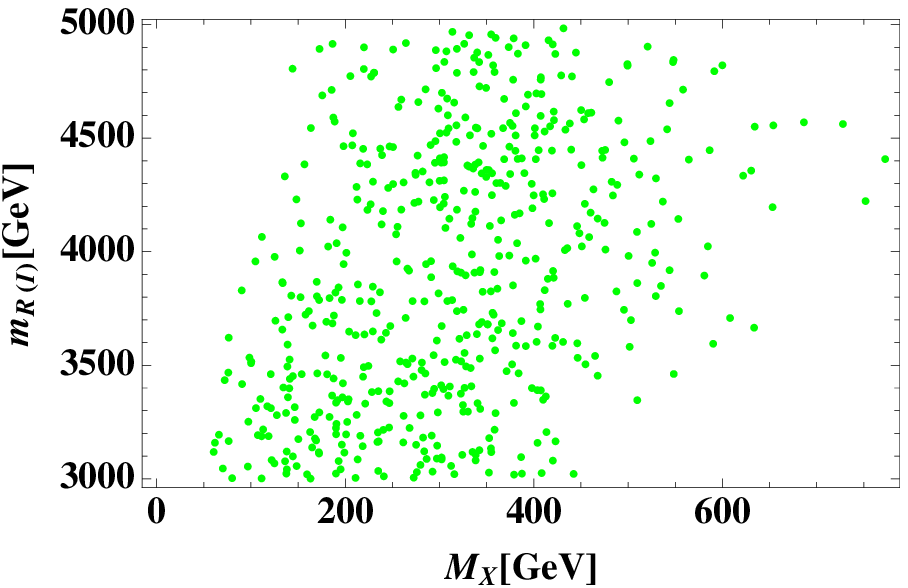} \qquad
\includegraphics[width=70mm]{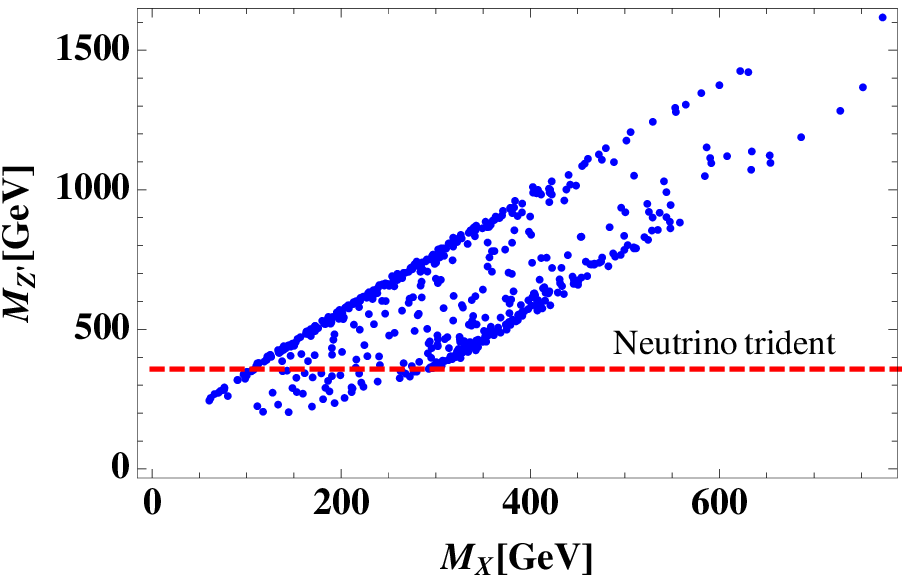}\qquad
\includegraphics[width=70mm]{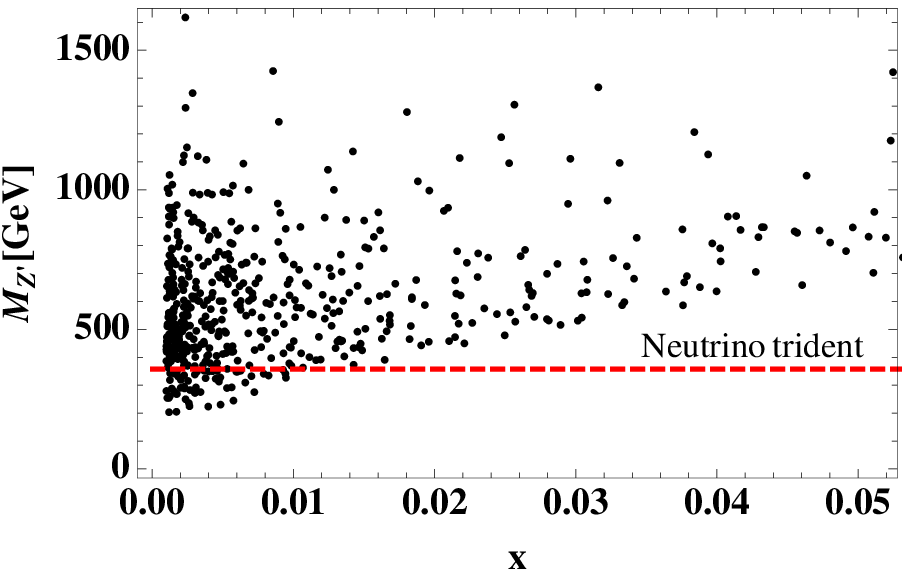}
\caption{
These three figures satisfy the neutrino oscillation data, LFVs, LEP bound, and thermal relic density of DM, where
the upper-left figure shows the scattering plot  between $M_X$ and $m_{R(I)}$,the upper-right figure shows the scattering plot  between $M_X$ and $M_{Z'}$, and the bottom figure shows the scattering plot  between $x$ and $M_{Z'}$. The red dashed line indicates the constraint from neutrino trident production for $g' = g_2 (\simeq 0.65)$} 
  \label{fig:DM-x-Zp}
\end{center}\end{figure}

The other properties are shown in Fig.~\ref{fig:DM-x-Zp} that satisfies the neutrino oscillation data, LFVs, 
LEP bound, and thermal relic density of DM 
where the allowed region of our DM mass is at $100-600$ [GeV], and the mass of $m_{R(I)}$ is likely to be 
a free parameter in the upper left figure. 
The correlation between $M_X$ and $M_{Z'}$ is shown in the right upper figure, in which the lower bound comes from the assumption $M_{Z'}< M_X$, while we took into account the constraints from LEP experiment in Eq.~(\ref{eq:lep}) and from neutrino trident production.
In addition, the bottom figure shows the soft correlation between $x$ and $M_{Z'}$, in which the upper bound also comes from the LEP experiment.

\section{ Conclusions and discussions}
In this paper, we have proposed a predictive radiative seesaw model at one-loop level with a flavor 
dependent gauge symmetry $U(1)_{xB_3-xe-\mu+\tau}$, in which we have considered the Majorana fermion dark 
matter. We have obtained a two zero texture with $A_1$ type that provides us the normal ordered neutrino mass 
spectra with $m_{\nu_3}\approx5\times 10^{-11}$ [GeV]. 
Also specific patterns of two Majorana phases are obtained in Fig.~\ref{fig:majo-p}. 
The other properties are shown in Fig.~\ref{fig:DM-x-Zp}, and we have found  the allowed region of our DM mass 
is at $100-600$ [GeV], and  the mass of $m_{R(I)}$ is likely to be the free parameter in the upper left figure. 
The correlation between $M_X$ and $M_{Z'}$ is shown in the right upper figure, in which the lower bound comes 
from the assumption $M_{Z'}< M_X$, and the constraints from LEP experiment and neutrino trident production 
are taken into account.
The bottom figure shows the soft correlation between $x$ and $M_{Z'}$, in which the upper bound of $x$ also comes from the LEP experiment for each value of $M_{Z'}$.

We also discussed phenomenology of $Z'$ which has flavor dependent couplings to SM fermions.
The flavor violating interaction in the down quark sector can induce a sizable contribution to the Wilson coefficient 
$C_9^{\mu \mu}$ which can be within $2 \sigma$ value obtained from global fitting by LHCb data. 
Although magnitude of our $|C_9^{\mu \mu}|$ is less than the best fit value it can be an explanation of anomalies 
in the measurements of $B \to K^* \mu^+ \mu^-$.
Remarkably we found anomaly in lepton-universality measurement 
${\rm BR}(B \to K \mu^+ \mu^-)/{\rm BR}(B \to K e^+ e^-)$ can be explained within our model. 
In addition, we estimated cross section of the process $pp \to Z' \to \mu^+ \mu^-$ at the LHC 13 TeV 
which provides a clear signature of flavor-dependent $Z'$ in our model.   
In particular, $Z'$ lighter than $O(1)$ TeV is allowed by current data and further parameter space can be tested 
in the future data of LHC experiments.


\section*{Acknowledgments}
\vspace{0.5cm}
H. O. is sincerely grateful for all the KIAS members, Korean cordial persons, foods, culture, weather, and all the other things.
This work is supported in part by National Research Foundation of Korea (NRF) Research Grant 
NRF-2015R1A2A1A05001869 (PK), and by the NRF grant funded by the Korea government (MSIP) 
(No. 2009-0083526) through Korea Neutrino Research Center at Seoul National University (PK). 


\end{document}